\documentclass[conference]{IEEEtran}
\usepackage{caption} 
\usepackage{lipsum}
\usepackage{indentfirst}
\usepackage{booktabs}
\usepackage{enumerate}
\usepackage{tabularx}
\usepackage{environ}         
\usepackage{etoolbox}        
\usepackage[dvipsnames]{xcolor}

\usepackage{cite}

\usepackage{graphicx}
\usepackage{wrapfig}
\usepackage[labelformat=simple]{subcaption}

\usepackage{float}
\usepackage{alltt}
\usepackage{boxedminipage}
\usepackage[pscoord]{eso-pic}

\newcommand{\placetextbox}[3]{
  \setbox0=\hbox{#3}
  \AddToShipoutPictureFG*{
    \put(\LenToUnit{#1\paperwidth},\LenToUnit{#2\paperheight}){\vtop{{\null}\makebox[0pt][c]{#3}}}%
  }%
}%

\usepackage{amsfonts, eqnarray, amsmath, amssymb}
\usepackage{mathtools}
\usepackage{xfrac}
\usepackage{breqn}
\usepackage{bm}

\usepackage{algorithm}
\usepackage{algpseudocode}
\usepackage{etoolbox}

\usepackage{array}

\usepackage{url}

\begin{document}
\title{Scalable~Coordinated~Learning~for~H2M/R Applications~over~Optical~Access~Networks~(Invited)}

\author{\IEEEauthorblockN{Sourav Mondal and Elaine Wong}
\IEEEauthorblockA{\textit{Department of Electrical and Electronic Engineering, The University of Melbourne, VIC 3010, Australia} \\
\texttt{sourav.mondal@unimelb.edu.au, ewon@unimelb.edu.au}}
}
\placetextbox{0.5}{0.06}{This article is accepted for publication in 29th Opto-Electronics and Communications Conference 2024 (OECC2024). Copyright @ IEEE.}%

\maketitle
\begin{abstract}
One of the primary research interests adhering to next-generation fiber-wireless access networks is human-to-machine/robot (H2M/R) collaborative communications facilitating Industry 5.0. This paper discusses scalable H2M/R communications across large geographical distances that also allow rapid onboarding of new machines/robots as $\sim72\%$ training time is saved through global-local coordinated learning.
\end{abstract}


\section{Introduction} \label{sec1}
In recent years, several inter-disciplinary technical paradigms like cyber-physical systems, Industrial IoT, robotics, big data, cloud/edge and cognitive computing, and virtual/augmented reality (VR/AR) have received significant attention from both industry and academia. The primary reason behind this development is the inclusion of industry vertical scenarios like Industry 4.0 in the fifth and beyond-fifth generation mobile technologies \cite{5g_iiot}. Although Industry 4.0 primarily involved connectivity among cyber-physical systems, Industry 5.0 will focus on the ``human and machine/robots/cobots" relationship \cite{Ind4_5} to ensure \emph{real-time monitoring} of products' condition, use, and the environment through sensors and external data sources, \emph{dynamic control} of product functions and personalized user experience through embedded software in the products, \emph{optimization} of use and performance of products, and \emph{autonomous delivery} of products through coordinated operations with other products and systems.\par
In Industry 5.0 scenarios, \emph{human operators (HOs)} will serve and maintain industrial plants from afar, as well as monitor, plan, and directly control the production processes in real-time or near-real-time via \emph{collaborator} robots/machines. The successful interaction between the HOs and machines/robots will rely on multi-modal information, e.g. audio, visual, and haptic data transmitted and received through state-of-the-art immersive devices e.g., smart glasses, smart suits, and VR/AR interfaces \cite{ewon}. For example, a HO is expected to use haptic gloves and/or suits with sensors and actuators to control robot(s) that mimic the actions of the HO. Successful H2M/R collaborations thus require a concerted effort in addressing several non-trivial technical challenges in haptics (acquisition, storage, transmission, and delivery of haptic data), intelligence (actions, movements, and feedback predictions), computation (process haptic feedback and run prediction algorithms), and communications (ensure low latency and high reliability) \cite{amin3}.\par
\begin{figure}[!t]
\centering
\includegraphics[width=0.97\columnwidth,keepaspectratio]{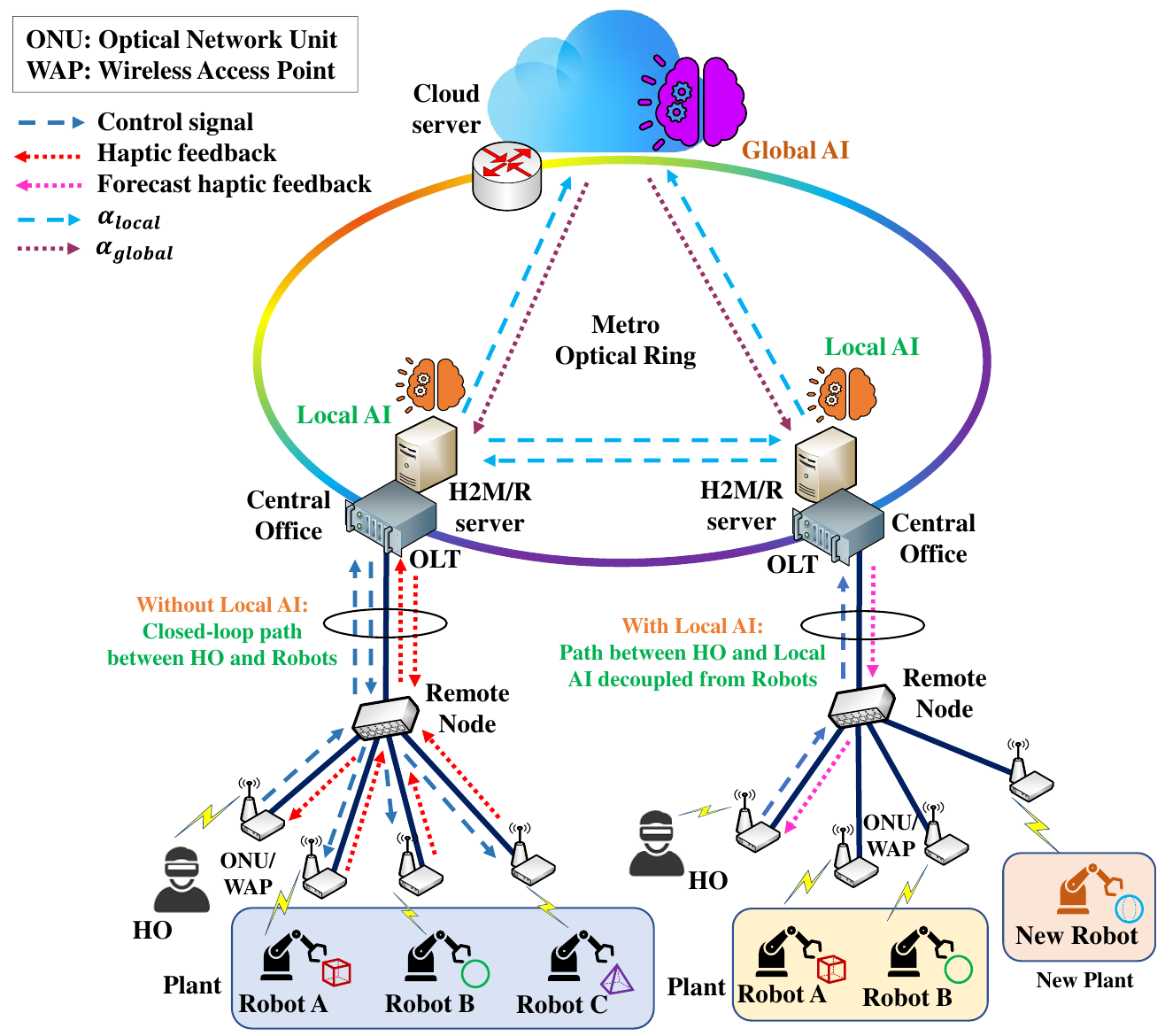}
\caption{Local AIs (edge servers) and Global AI (cloud server) supporting H2M/R collaboration over fiber-wireless access networks.}
\label{fed_arc}
\end{figure}
\setlength{\textfloatsep}{5pt}
In particular, the perception of real-time and/or near real-time H2M/R collaboration require millisecond \cite{NTTWhite} round-trip latency between the HO and machine/robot. This becomes extremely challenging for geographically separated HO-machine/robot pairs whereby transmissions may need to traverse beyond the wireless access segment and through to optical access and even metro access segments. With organizations establishing multi-campus or multi-site presence, the flexibility of supporting large-scale and scalable H2M/R collaborations across large geographical distances is, however, still an open challenge \cite{ml_tute}. Drawing from \cite{sourav_hpt}, multiple AI-embedded edge servers termed \emph{Local AIs} could potentially be deployed, each serving a HO and multiple machine/robots. Even then, when multiple Local AIs serving HO-machine/robot pairs are engaged in identical applications and are interacting with similar remote machines/robots, the time and computational costs in training multiple Local AIs can become inefficient. Furthermore, newly-introduced machines/robots will not only require re-training of the corresponding Local AI to ensure haptic feedback forecast accuracy is maintained, but the same re-training will need to be performed every time a similar object is introduced elsewhere in the network. Moreover, when retraining is performed at a Local AI, this will disrupt the operation of other machine/robots connected to that Local AI. We thus proposed the Global-Local AI coorDinated learning (GLAD) framework in \cite{sm_glad} to support the dynamic addition and rapid onboarding of new machines/robots. In this invited article, we critically review GLAD and discuss how it empowers Local AIs to accurately forecast haptic feedback of newly-introduced machine/robots by building upon the generalized federated learning protocol in \cite{Google} to enable Local AIs to collaboratively learn from each other and also through a remote cloud server.\par

\section{The GLAD Framework} \label{sec2}
In this article, H2M/R collaborations are exemplified through geographically distributed industrial plants that are connected via a converged fiber-wireless access network architecture, as shown in Fig. \ref{fed_arc}. Each HO and its associated machine/robots are optically and wirelessly connected via a Central Office (CO) that houses a Local AI. The HO is expected to immerse in the remote industrial plants of the machine/robots it controls. In operation, the HO controls a machine/robot by sending control signals to the machine/robot which mimics the HO's gestures. These signals typically constitute flexible and orientation sensors for mapping the coordinates, orientations, and pressure on fingers or involved body parts of the HO.\par
Without Local AI, the exchange of control signals and haptic feedback samples between HO and machine/robots creates a closed-loop where round-trip latency is affected by the distance and network load. This is depicted by the left fiber-wireless access network in Fig. \ref{fed_arc}. Further, the need to meet stringent latency requirements to realize real-time or near-real-time H2M/R thus limits the deployable HO-machine/robot distance. The introduction of a Local AI through the GLAD framework will minimize the impact of network load and HO-machine/robot distance on the round-trip latency. That is, the Local AI will forecast and expedite the transmission of haptic feedback to the HO, thereby reducing the round-trip latency and allowing the HO-machine/robot distance to be extended, as illustrated in the right fiber-wireless access network.\par
The Local AI uses a binary classifier module on the control signals to predict if the HO is going to touch any machine/robot and a reinforcement learning module to forecast the haptic feedback samples to the HO. Furthermore, the proposed GLAD framework facilitates the introduction of new machine/robots without operational interruption to existing machine/robots. The Local AIs, each serving a HO and one or more machine/robots, collaboratively learn from each other as well as through a remote cloud server forming the Global AI module.

\section{Performance Evaluation} \label{sec3}
To train and validate the GLAD framework, H2M/R collaboration through a haptic VR experiment is established to collect control and haptic feedback data between a HO and a virtual machine/robot handling virtual objects of differing shapes, material, textures, and sizes \cite{sm_glad}. In the experiment, the HO uses VR haptic gloves and VR goggles to immerse in the virtual scene. Thus, any HO hand movement is mimicked by the virtual hands. Upon touching any virtual object, haptic feedback is felt by the HO via the VR haptic gloves. In evaluating GLAD, a dataset of 12000 (70\% training, 30\% validation) samples is used to classify ``touch/no-touch" events while a separate dataset of 4000 samples are used for profiling the forecast haptic feedback of each object at the Local AI.\par
\begin{figure}[!t]
\centering
\includegraphics[width=0.8\columnwidth]{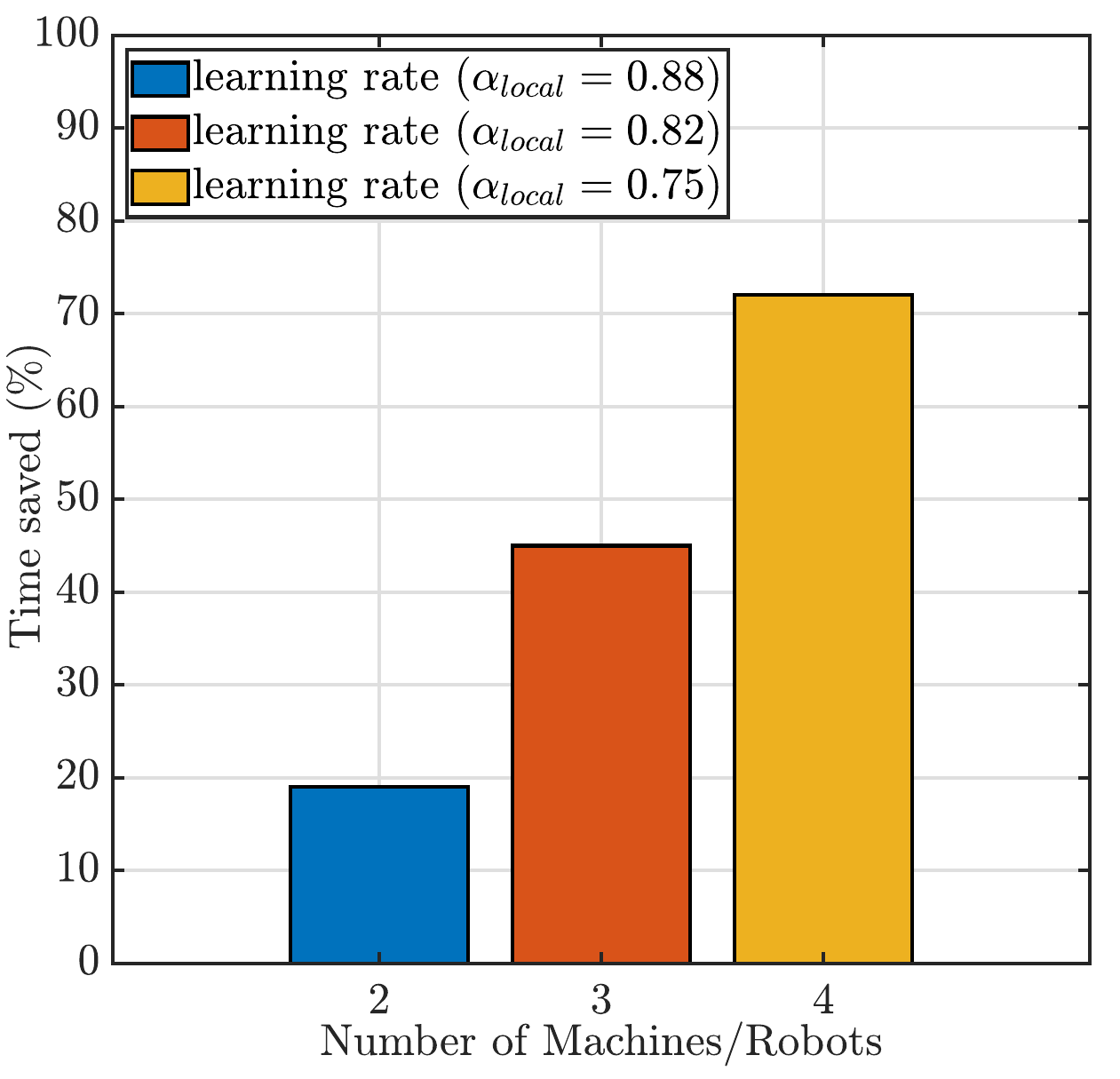}
\caption{Percentage of Local AI training time saved by GLAD against total number of machine/robots.}
\label{fed_acc}
\end{figure}
\setlength{\textfloatsep}{5pt}
The traffic distribution of the control and haptic feedback signals are investigated by performing three different test scenarios, namely grabbing a virtual rubber ball, wooden cube, and circular cube, respectively. Histograms from the timestamps of the control signals and haptic feedback for all three test scenarios were evaluated, with results fitting a \emph{generalized Pareto distribution} with a 5\% significance level \cite{sourav_hpt}. A significant benefit of GLAD is to ensure all Local AIs will dynamically support the rapid onboarding of newly-introduced machine/robots with high haptic feedback forecast accuracy and without operational disruption to existing machine/robots.\par
\begin{figure*}[!t]
  \centering
  \subfloat[Human operator$\,\to\,$machine/robot closed loop]{%
    \includegraphics[width=0.9\columnwidth]{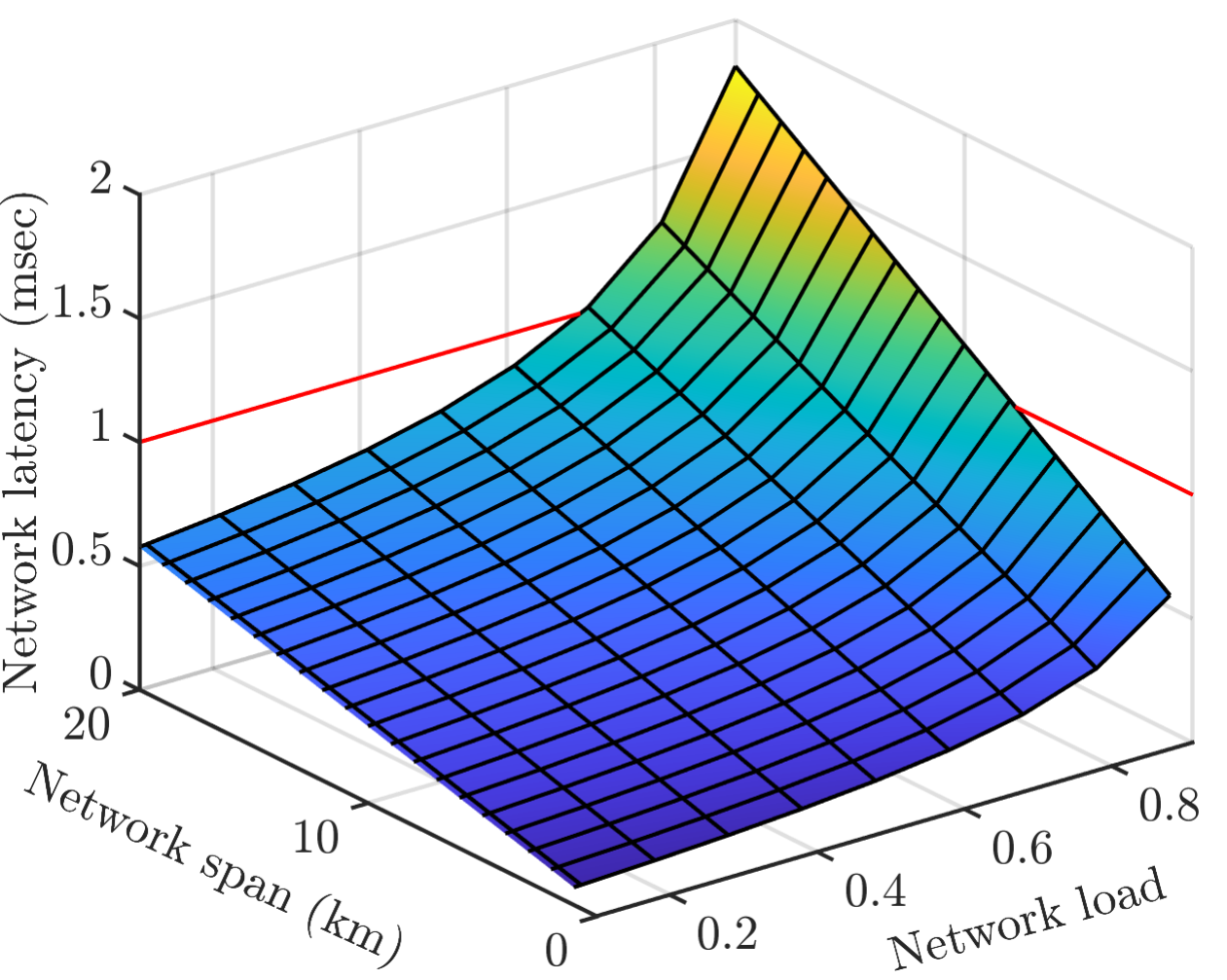}\label{lat1}%
  }
  \subfloat[(Human operator$\,\to\,$Local AI) $+$ (Local AI$\,\to\,$Human operator)]{%
   \includegraphics[width=0.9\columnwidth]{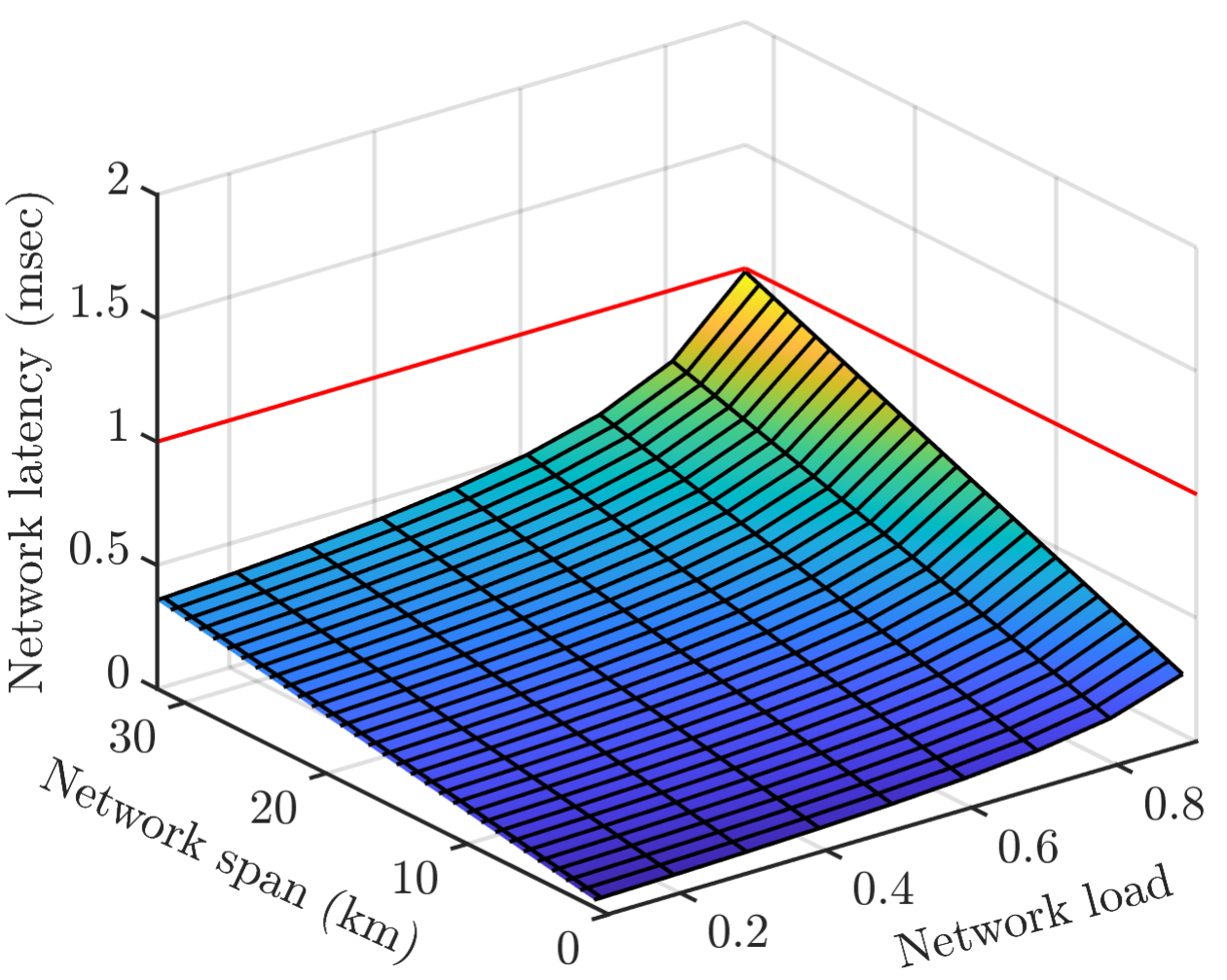}\label{lat3}%
  }

  \caption{Transmission latencies between human operator, Local AI, and machine/robot over an XG-PON with a 1:16 split ratio.}
  \label{ms_dst}
\end{figure*}
Starting from one machine/robot, we investigate the cumulative accuracy in haptic feedback forecasting as a function of iteration in which a new machine/robot is added periodically after some iterations. In \cite{sm_glad}, we showed that the accuracy decreases from 100\% (with 1 machine/robot present) with every new machine/robot introduced within a Local AI. The value of learning rate parameter $\alpha_{local}$ that achieves the highest accuracy for a Local AI primarily depends on the number of machine/robots $M$ and the correlation among the haptic feedback samples $\tau$. Hence, whenever a new machine/robot is introduced within the Local AI, a new optimal value of $\alpha_{local}$ is required to be obtained as exploration and exploitation may happen with unknown rewards.\par 
Through numerical calculations of an XG-PON with a split ratio of 1:16, in Fig. \ref{lat1} we plot the round-trip transmission latency between a HO and a machine/robot. This comprises the time it takes the control signal from the HO to reach the machine/robot and for the haptic feedback from the machine/robot to reach the HO. For a typical 20 km XG-PON span, i.e. 40 km HO-machine/robot distance, results show latencies exceeding 1 ms under high network load. Nonetheless, when haptic feedback samples are forecasted using the Local AI through GLAD, the transmission latency of control signals from the HO to Local AI and to the machine/robot reduces drastically. Fig. \ref{lat3} plots the round-trip transmission latency comprising the control signal from the HO to AI and the \emph{forecast haptic feedback} from the Local AI to the HO. Clearly, the capability to forecast haptic feedback samples at the Local AI enables the 1 ms latency requirement to be met under an increased HO-machine/robot distance. As a consequence, a network span much larger than 30 km, i.e. beyond 60 km HO-machine/robot distance, can be achieved especially at low to mid network load \cite{sourav_hpt}. \par

\section{Conclusions} \label{sec4}
In this invited article, we presented a Global-Local AI coordinated learning (GLAD) framework over converged wireless-optical access networks that allows rapid onboarding of machines/robots without disrupting ongoing operations and at extended human-machine/robot distances. Numerical calculations of an XG-PON implementing GLAD are shown to be able to support H2M/R collaborations over 60 km under a network load of 0.8, and potentially further than 60 km under low to mid network load.

\bibliographystyle{IEEEtran}
\bibliography{IEEEabrv,references}

\end{document}